\documentclass[preprintnumbers,amsmath,amssymb]{revtex4}

\usepackage{graphicx}
\usepackage{dcolumn}
\usepackage{bm}

\begin{document}

\begin{center}
{\Large \bf Probabilistically Cloning and Quantum Computation $^*$}\\[0.3cm]

GAO Ting$^{1,2}$, YAN  Feng-Li$^{3,4}$, Wang Zhi-Xi$^2$\\[0.1cm]

 {\footnotesize $^1$\sl College of Mathematics and Information Science, Hebei Normal University, Shijiazhuang 050016, China\\
 $^2$ Department of Mathematics, Capital Normal University, Beijing 100037, China\\
$^3$ CCAST (World Laboratory), P.O. Box 8730, Beijing 100080, China\\
$^4$ College of Physics, Hebei Normal University, Shijiazhuang 050016, China}\\[0.2cm]

\begin{minipage} {16cm}
{\noindent\footnotesize \sl We discuss the usefulness of quantum cloning and present examples of quantum
computation tasks for which cloning offers an advantage which cannot be matched by any approach that does not
resort to it. In these quantum computations, we need to distribute quantum information contained in states about
which we have some partial information. To perform quantum computations, we use state-dependent probabilistic
quantum cloning procedure  to distribute quantum information in the middle
of a quantum computation.}\\

PACS: 03.67.Lx, 42.50.Dv
\end{minipage}
\end{center}

Cloning is very useful  in classical computing and easy  to accomplish  with classical information. However,
quantum cloning turns out not to be possible in general in quantum mechanics. This no-cloning theorem,
independently  discovered by Wootters and Zurek$^{[1]}$ and Dieks$^{[2]}$ in the early 1980s, is one of the most
fundamental differences between classical and quantum information theories. It tells us that an unknown quantum
state can not be copied exactly. Since determinately perfect copying is impossible, much effort has been put
into developing optimal cloning processes.$^{[3-14]}$
  The universal quantum cloning machines
 were first invented by Bu$\check{\rm z}$ek and Hillery$^{[3]}$ and developed by other authors.$^{[4-12]}$
 The  another kind of cloning procedure first designed by  Duan and Guo$^{[13, 14]}$ is nondeterministic,
consisting in adding an ancilla, performing unitary operations and measurements, with a postselection of the
measurement results. The resulting clones are perfect, but the procedure only succeeds with a certain
probability $p<1$. The imperfect nature of quantum cloning procedure results in lower chances of getting the
correct computational outputs at the end. Nonetheless, in some cases cloning improves our chances of correctly
computing.
 Up to now, only a few examples are available to show that quantum cloning is
useful in quantum computation.$^{[15-18]}$ In this letter  we investigate  the possible use of quantum cloning
machine and present examples of quantum computation tasks for which cloning offers an advantage which cannot be
matched by any approach that does not resort to quantum cloning. In these quantum computations, we need to
distribute quantum information contained in states about which we have some partial information. To perform
quantum computations, we use state-dependent probabilistic quantum cloning procedure discussed by Duan and
Guo$^{[13, 14]}$ to distribute quantum information in the middle of a quantum computation.
 Next we propose a generalization of  examples.$^{[15, 16, 17]}$

Let us consider the scenario. There are two different quantum computations with first computational step $U_0$
in common. We need to find a scheme that obtains the two computation results with as large a probability as
possible, using $U_0$ only once (this may happen if $U_0$ is a complex, lengthy computation). Possible schemes
may or may not resort to cloning to distribute quantum information; the quantum computational tasks below show
that performance is enhanced if we distribute quantum information using quantum cloning.

  Suppose that there are three quantum black-boxes. What each
black-box  does is to accept one $2$-level quantum system as an input and apply a unitary operator  to it,
producing the evolved state as an output. The black-boxes  consist of arbitrary quantum circuits that query a
given function only once. The query of function $f_i$ is the unitary such that
$|x\rangle|y\rangle\rightarrow|x\rangle|y\oplus f_i(x)\rangle$, where the symbol $\oplus$
 denotes the bitwise $XOR$ operation. Our task will involve determining two functionals, one depending only on
 $f_0$ and $f_1$, and the other on $f_0$ and $f_2$.

 Consider all functions $h_i$ which take $n$ bits to one bit  and   write  $h_{a_0a_1\cdots
a_{2^n-1}}$ to stand for the function $h$ such that $h(k)=a_k$, $k=0, 1, \cdots, 2^n-1$ ( e.g.
$h_{a_0a_1a_2a_3a_4a_5a_6a_7}$ denotes $h(000)=a_0$, $h(001)=a_1$, $h(010)=a_2$, $h(011)=a_3$, $h(100)=a_4$,
$h(101)=a_5$, $h(110)=a_6$, $h(111)=a_7$ ).  We define some sets of functions that will be useful in stating our
task:

\[S^3_{f_0}=\{h_{01000000}, h_{01010101},
h_{11000011}\},\]
{\noindent --------------------------------------------  \\
{\footnotesize $^*$Supported by National Natural Science Foundation of China under Grant No. 10271081 and  Hebei
Natural Science Foundation under Grant No. A2004000141.}}
\newpage
{\parindent=0cm
\begin{eqnarray*} &&S^3_1=\{h_{01000000}, h_{10110000}, h_{10001100}, h_{00100110},
h_{00010101}, h_{10000011}, h_{00101001}, h_{00011010}\},\nonumber\\
&&S^3_2=\{h_{00000000}, h_{00001111}, h_{01010101}, h_{00110011}, h_{10011001}, h_{11000011}, h_{01101001},
h_{10100101}\},
\end{eqnarray*}
\[S^3_{f_{12}}=S^3_1\cup S^3_2,\]}
\begin{eqnarray*}
&&H^3_{00000000}=\{h_{00000000}, h_{11111111}\}, H^3_{00001111}=\{h_{00001111}, h_{11110000}\},
H^3_{01010101}=\{h_{01010101}, h_{10101010}\},\\
&&H^3_{00110011}=\{h_{00110011}, h_{11001100}\}, H^3_{10011001}=\{h_{10011001}, h_{01100110}\},
H^3_{11000011}=\{h_{11000011}, h_{00111100}\},\\
&&H^3_{01101001}=\{h_{01101001}, h_{10010110}\}, H^3_{10100101}=\{h_{10100101}, h_{01011010}\},
\end{eqnarray*}
\begin{eqnarray*}
S^3_f=H^3_{00000000}\cup H^3_{00001111}\cup H^3_{01010101}\cup H^3_{00110011} \cup H^3_{10011001}\cup
H^3_{11000011}\cup H^3_{01101001}\cup H^3_{10100101}.
\end{eqnarray*}

\[S^{n+1}_{f_0}=\{h_{a_0a_1a_2\cdots a_{2^n-1}a_0a_1a_2\cdots a_{2^n-1}} ~ | ~~ h_{a_0a_1a_2\cdots
a_{2^n-1}}\in S^n_{f_0}\},\]
\begin{eqnarray*}
S^{n+1}_1=\{h_{a_0a_1a_2\cdots a_{2^n-1}a_0a_1a_2\cdots a_{2^n-1}},
h_{\overline{a_0}\overline{a_1}\overline{a_2}\cdots
\overline{a_{2^n-1}}a_0a_1a_2\cdots a_{2^n-1}} ~ | ~~ h_{a_0a_1a_2\cdots a_{2^n-1}}\in S^n_1\},\\
S^{n+1}_2=\{h_{a_0a_1a_2\cdots a_{2^n-1}a_0a_1a_2\cdots a_{2^n-1}},
h_{\overline{a_0}\overline{a_1}\overline{a_2}\cdots \overline{a_{2^n-1}}a_0a_1a_2\cdots a_{2^n-1}} ~ | ~~
h_{a_0a_1a_2\cdots a_{2^n-1}}\in S^n_2\},
\end{eqnarray*}
\[S^{n+1}_{f_{12}}=S^{n+1}_1\cup S^{n+1}_2,\]
\begin{eqnarray*}
H^{n+1}_{a_0a_1a_2\cdots a_{2^n-1}a_0a_1a_2\cdots a_{2^n-1}}=\{h_{a_0a_1a_2\cdots a_{2^n-1}a_0a_1a_2\cdots
a_{2^n-1}}, h_{\overline{a_0}\overline{a_1}\overline{a_2}\cdots
\overline{a_{2^n-1}}\overline{a_0}\overline{a_1}\overline{a_2}\cdots \overline{a_{2^n-1}}} ~ |
~~ h_{a_0a_1a_2\cdots a_{2^n-1}}\in S^n_2 \},\\
H^{n+1}_{\overline{a_0}\overline{a_1}\overline{a_2}\cdots \overline{a_{2^n-1}}a_0a_1a_2\cdots
a_{2^n-1}}=\{h_{\overline{a_0}\overline{a_1}\overline{a_2}\cdots \overline{a_{2^n-1}}a_0a_1a_2\cdots a_{2^n-1}},
h_{a_0a_1a_2\cdots a_{2^n-1}\overline{a_0}\overline{a_1}\overline{a_2}\cdots \overline{a_{2^n-1}}} ~ | ~~
h_{a_0a_1a_2\cdots a_{2^n-1}}\in S^n_2\},
\end{eqnarray*}
\begin{eqnarray*}S^{n+1}_f=\bigcup_{h_{a_0a_1a_2\cdots a_{2^n-1}}\in S^n_2} (H^{n+1}_{a_0a_1a_2\cdots a_{2^n-1}a_0a_1a_2\cdots a_{2^n-1}}\cup
H^{n+1}_{\overline{a_0}\overline{a_1}\overline{a_2}\cdots \overline{a_{2^n-1}}a_0a_1a_2\cdots a_{2^n-1}}).
\end{eqnarray*}
Here $n= 3, 4, \cdots $, $\overline{a_k}=\biggl\{\begin{array}{ll}
  0, & \text{if }  a_k=1 \\
  1, & \text{if }  a_k=0,
\end{array}$
$k=0, 1, 2, \cdots, 2^n-1$.

Now we  randomly choose  a function $f_0\in S^n_{f_0}$ and then pick two other functions $f_1$ and $f_2$ from
the set $S^n_{f_{12}}$, also at  random  but satisfying:
\begin{equation}\label{f123}
f_0\oplus f_1, ~~f_0\oplus f_2\in S^n_f.
\end{equation}
Here the  symbol $\oplus$  is addition modulo 2.  The task will be to find in which of the $2^n$ sets $H^n$'s
lie each of the functions $f_0\oplus f_1$ and $f_0\oplus f_2$, applying quantum circuits that query $f_0$,
$f_1$, and $f_2$ at most once each.  Our score will be given by the average probability of successfully guessing
both correctly.

 Just as [15] the best no-cloning strategy is
as follows. First, from  Eq.(\ref{f123}) we  know that  both $f_1$ and $f_2$ must be in $S^n_1$ if
$f_0=h_{\{01000000\}^{2^{n-3}}}$, and  $f_1$ and $f_2$ must belong to $S^n_2$ if $f_0$ is either
$h_{\{01010101\}^{2^{n-3}}}$ or $h_{\{11000011\}^{2^{n-3}}}$, where ${\{a_0a_1\cdots a_7\}^{2^{n-3}}}$ means
$2^{n-3}$ copies of $a_0a_1\cdots a_7$. It implies that
 the probability of  both $f_1$ and $f_2$  in $S^n_2$  is $2/3$. Assume that it is the case, then
we can discriminate between the two possibilities for $f_0$ with a single, classical function call. Furthermore,
by using the quantum circuit in Fig.1 twice (once each with $f_1$ and $f_2$) we can distinguish the $2^n$
possibilities for functions $f_1$ and $f_2$, because  this quantum circuit results in one of the $2^n$
orthogonal states
\begin{equation}
|\varphi_i\rangle=2^{-\frac{n}{2}}\sum_{x=0}^{2^n-1}(-1)^{f_i(x)}|x\rangle.
 \end{equation}
Thus we can determine functions $f_0$, $f_1$, and $f_2$ correctly with probability $p=2/3$ and accomplish our
task. Even in the case where the initial assumption about $f_0$ was wrong,  the chances of guessing right  are
$1/2^{2n}$. Therefore, the best no-cloning average score is
\begin{equation}
p_1=\frac {2}{3} + \frac {1}{3}\times \frac{1}{2^{2n}} < 0.7.
\end{equation}

  Next we will see that  the task which can be much  better performed if we use  probabilistic quantum cloning.
   The quantum circuit that we use to solve this problem is depicted in Fig.~2.

Immediately after querying function $f_0$, we obtain one of three possible linearly independent states (each
corresponding to one of the possible $f_0$'s): {\footnotesize\begin{eqnarray}
|h_{\{01000000\}^{2^{n-3}}}\rangle\equiv &
2^{-\frac{n}{2}}\sum^{2^{n-3}-1}_{j=0}(|2^3j\rangle-|2^3j+1\rangle+|2^3j+2\rangle+|2^3j+3\rangle+|2^3j+4\rangle+|2^3j+5\rangle
+|2^3j+6\rangle+|2^3j+7\rangle),\label{h11}\\
|h_{\{01010101\}^{2^{n-3}}}\rangle\equiv &
2^{-\frac{n}{2}}\sum^{2^{n-3}-1}_{j=0}(|2^3j\rangle-|2^3j+1\rangle+|2^3j+2\rangle-|2^3j+3\rangle+|2^3j+4\rangle-|2^3j+5\rangle
+|2^3j+6\rangle-|2^3j+7\rangle),\label{h12}\\
|h_{\{11000011\}^{2^{n-3}}}\rangle\equiv &
2^{-\frac{n}{2}}\sum^{2^{n-3}-1}_{j=0}(-|2^3j\rangle-|2^3j+1\rangle+|2^3j+2\rangle+|2^3j+3\rangle+|2^3j+4\rangle+|2^3j+5\rangle
-|2^3j+6\rangle-|2^3j+7\rangle).\label{h13}
\end{eqnarray}}

  By Theorem 2 in [13] and
the method in  [16] we derive the following exact achievable cloning efficiencies (defined as the probability of
cloning successfully) for each of states (\ref{h11})--(\ref{h13})
\begin{eqnarray}
 &\gamma_1\equiv \gamma(|h_{\{01000000\}^{2^{n-3}}}\rangle)=\frac {7}{127},\label{r11}\\
 &\gamma_2\equiv \gamma(|h_{\{01010101\}^{2^{n-3}}}\rangle)=\gamma_3\equiv
 \gamma(|h_{\{11000011\}^{2^{n-3}}})=\frac {112}{127}.\label{r123}
\end{eqnarray}

After the cloning process the measurement  outcome on a "flag" subsystem  will tell us whether the cloning was
successful or not. For this particular cloning process, the probability of success is, on average, $P_{\rm
success}=(\gamma_1+\gamma_2+\gamma_3)/3=\frac {77}{127}$. If it was successful, then each of the cloning
branches goes through the second part of the circuit in Fig.2, to yield one of the $2^n$ orthogonal states:
\begin{eqnarray}
  |h_{a_0a_1a_2\cdots a_{2^n-1}}\rangle &=&2^{-\frac {n}{2}}\sum_{x=0}^{2^n-1}(-1)^{a_x}|x\rangle,\label{h}
\end{eqnarray}
which can be discriminated unambiguously. Here $h_{a_0a_1a_2\cdots a_{2^n-1}}\in S^n_2$.   Therefore, if the
cloning process is successful, we manage to accomplish our task.

However, the cloning process may fail with  probability $(1-P_{\rm success})$. If this occurs, it is more likely
to be $h_{\{01000000\}^{2^{n-3}}}$ than the other two, because of  the relatively low cloning efficiency for the
state in Eq.(\ref{h11}), in relation to the states in Eqs.(\ref{h12}) and (\ref{h13}) [see Eqs. (\ref{r11}) and
(\ref{r123})]. Guessing that $f_0=h_{\{01000000\}^{2^{n-3}}}$, we are right with probability
\begin{equation}
p_{\{01000000\}^{2^{n-3}}}=\frac {(1-\gamma_1)}{(1-\gamma_1)+(1-\gamma_2)+(1-\gamma_3)}=\frac {4}{5}.
\end{equation}
Note that only the $2^n$ functions in $S^n_1$ can be candidates for $f_1$ and $f_2$ on condition that
$f_0=h_{\{01000000\}^{2^{n-3}}}$. These $2^n$ possibilities can be  discriminated unambiguously by run a circuit
like that of Fig.1 twice, once with $f_1$ and once with $f_2$, since the circuit produces one of $2^n$
orthogonal states, each corresponding to one of the $2^n$ possibilities for $f_i$. Therefore if our  guess that
$f_0=h_{\{01000000\}^{2^{n-3}}}$ was correct, we are able to find the correct $f_1$ and $f_2$ and therefore
accomplish our task. In the case that $f_0\neq h_{\{01000000\}^{2^{n-3}}}$ after all, we may still have guessed
the right sets  with probability $1/2^{2n}$.

The above considerations lead to an overall probability of success given by
\begin{eqnarray}\label{y}
p_2&=&p_{\rm success}+(1-p_{\rm success})[p_{\{01000000\}^{2^{n-3}}}+(1-p_{\{01000000\}^{2^{n-3}}})\frac{1}{2^{2n}}]\nonumber\\
   &=&\frac{117}{127}+\frac{5}{127\times 2^{2n-1}}\nonumber\\
   &>&\frac{117}{127}\nonumber\\
   &>&p_1.
\end{eqnarray}
 It shows
that this cloning approach is more efficient than the previous one, which does not use cloning.\\

If we take
\[S^2_{f_0}=\{h_{0100}, h_{0011},
h_{1001}\},\]
\begin{eqnarray*}
S^2_1=\{h_{0001}, h_{0010}, h_{0100}, h_{1000}\}, S^2_2=\{h_{0000},  h_{0011}, h_{0101},  h_{1001}\},
\end{eqnarray*}
\[S^2_{f_{12}}=S^2_1\cup S^2_2,\]
\begin{eqnarray*}
H^2_{0000}=\{h_{0000}, h_{1111}\}, H^2_{0101}=\{h_{0101}, h_{1010}\}, H^2_{0011}=\{h_{0011}, h_{1100}\},
H^2_{1001}=\{h_{1001}, h_{0110}\},
\end{eqnarray*}
\begin{eqnarray*}S^2_f=H^2_{0000}\cup H^2_{0101}\cup
H^2_{0011} \cup H^2_{1001};\end{eqnarray*}

\[S^{n+1}_{f_0}=\{h_{a_0a_1a_2\cdots a_{2^n-1}a_0a_1a_2\cdots a_{2^n-1}} ~ | ~~ h_{a_0a_1a_2\cdots
a_{2^n-1}}\in S^n_{f_0}\},\]
\begin{eqnarray*}
S^{n+1}_1=\{h_{a_0a_1a_2\cdots a_{2^n-1}a_0a_1a_2\cdots a_{2^n-1}},
h_{\overline{a_0}\overline{a_1}\overline{a_2}\cdots
\overline{a_{2^n-1}}a_0a_1a_2\cdots a_{2^n-1}} ~ | ~~ h_{a_0a_1a_2\cdots a_{2^n-1}}\in S^n_1\},\\
S^{n+1}_2=\{h_{a_0a_1a_2\cdots a_{2^n-1}a_0a_1a_2\cdots a_{2^n-1}},
h_{\overline{a_0}\overline{a_1}\overline{a_2}\cdots \overline{a_{2^n-1}}a_0a_1a_2\cdots a_{2^n-1}} ~ | ~~
h_{a_0a_1a_2\cdots a_{2^n-1}}\in S^n_2\},
\end{eqnarray*}
\[S^{n+1}_{f_{12}}=S^{n+1}_1\cup S^{n+1}_2,\]
\begin{eqnarray*}
H^{n+1}_{a_0a_1a_2\cdots a_{2^n-1}a_0a_1a_2\cdots a_{2^n-1}}=\{h_{a_0a_1a_2\cdots a_{2^n-1}a_0a_1a_2\cdots
a_{2^n-1}}, h_{\overline{a_0}\overline{a_1}\overline{a_2}\cdots
\overline{a_{2^n-1}}\overline{a_0}\overline{a_1}\overline{a_2}\cdots \overline{a_{2^n-1}}} ~ |
~~ h_{a_0a_1a_2\cdots a_{2^n-1}}\in S^n_2 \},\\
H^{n+1}_{\overline{a_0}\overline{a_1}\overline{a_2}\cdots \overline{a_{2^n-1}}a_0a_1a_2\cdots
a_{2^n-1}}=\{h_{\overline{a_0}\overline{a_1}\overline{a_2}\cdots \overline{a_{2^n-1}}a_0a_1a_2\cdots a_{2^n-1}},
h_{a_0a_1a_2\cdots a_{2^n-1}\overline{a_0}\overline{a_1}\overline{a_2}\cdots \overline{a_{2^n-1}}} ~ | ~~
h_{a_0a_1a_2\cdots a_{2^n-1}}\in S^n_2\},
\end{eqnarray*}
\begin{eqnarray*}S^{n+1}_f=\bigcup_{h_{a_0a_1a_2\cdots a_{2^n-1}}\in S^n_2} (H^{n+1}_{a_0a_1a_2\cdots a_{2^n-1}a_0a_1a_2\cdots a_{2^n-1}}\cup
H^{n+1}_{\overline{a_0}\overline{a_1}\overline{a_2}\cdots \overline{a_{2^n-1}}a_0a_1a_2\cdots a_{2^n-1}}).
\end{eqnarray*}
Here $n=2, 3, 4, \cdots $, $\overline{a_k}=\biggl\{\begin{array}{ll}
  0, & \text{if }  a_k=1 \\
  1, & \text{if }  a_k=0,
\end{array}$
$k=0, 1, 2, \cdots, 2^n-1$. Following the same procedure as before, we derive that the best no-cloning average
score is still
\begin{equation}
p_1=\frac {2}{3} + \frac {1}{3}\times \frac{1}{2^{2n}} < 0.7,
\end{equation}\\[0.2cm]
but the analytic achievable   cloning efficiencies are
\begin{eqnarray}
 &\gamma_1\equiv \gamma(|h_{\{0100\}^{2^{n-2}}}\rangle)=\frac {1}{7},\\
 &\gamma_2\equiv \gamma(|h_{\{0011\}^{2^{n-2}}}\rangle)=\gamma_3\equiv
 \gamma(|h_{\{1001\}^{2^{n-2}}})=\frac {4}{7},
\end{eqnarray}
and the overall probability of success with cloning is
\begin{eqnarray}
p_2&=&\frac{5}{7}+\frac{2}{7\times 2^{2n-1}}\nonumber\\
   &>&\frac{5}{7}\nonumber\\
   &>&p_1.
\end{eqnarray}

Besides the larger probability of obtaining the correct result, the quantum cloning offers another advantage:
 the measurement of the 'flag' state makes us to be confident about having the correct result in a larger
 fraction of our attempts. For the above second probabilistic cloning machine described by $\gamma_1=\frac {1}{7}$,
 $\gamma_2=\gamma_3=\frac {4}{7}$  this fraction was
 $\frac{3}{7}$, but this can be improved by choosing different cloning machine, e.g. the first one above
 characterized by $\gamma_1=\frac {7}{127}$,
 $\gamma_2=\gamma_3=\frac {112}{127}$,  the fraction of which is $\frac{77}{127}$.

In summary, we study the possible use of quantum cloning machine and give examples of quantum computation tasks
which attain optimal performance by using an intermediate quantum cloning step. We show that preserving and
distributing quantum information through a cloning process offers advantages over extracting classical
information mid-way in the quantum
computation. \\[0.2cm]

{\parindent=0cm \Large\bf References \\~~}

{\footnotesize\noindent
$~[1]$    Wootters W K and  Zurek W H 1982 {\it Nature (London)} {\bf 299} 802\\
$~~[2]$  Dieks D 1982 {\it Phys. Lett.} A {\bf 92} 271\\
$~~[3]$   Bu\v{z}ek V and  Hillery M 1996 {\it Phys. Rev.} A {\bf 54} 1844\\
$~~[4]$   Gisin N and  Massar S 1997 {\it Phys. Rev. Lett.} {\bf 79} 2153 \\
$~~[5]$  Gisin N 1998 {\it Phys.Lett.} A {\bf 242} 1\\
$~~[6]$  Werner R F 1998 {\it Phys. Rev.} A {\bf 58} 1827\\
$~~[7]$  Keyl M and  Werner R F 1999 {\it J. Math. Phys.} {\bf 40} 3283\\
$~~[8]$  Bru${\ss}$ D and  Macchiavello C 1999 {\it Phys. Lett.} A {\bf 253} 249\\
$~~[9]$  Bru${\ss}$ D,  Ekert A and  Macchiavello C 1998  {\it Phys. Rev. Lett.} {\bf 81} 2598\\
$[10]$ Bu{\v{z}}ek V,  Hillery M and  Bednik R 1998 {\it Acta Phys. Slov.} {\bf 48} 177\\
$[11]$  Nicolas J. Cerf 2000 {\it J. Mod. Opt.} {\bf 47} 187\\
$[12]$   Bu{\v{z}}ek V,  Braunstein S,  Hillery M and  Bru$\ss$ D 1997 {\it Phys. Rev.} A {\bf 56} 3446\\
$[13]$   Duan L M and  Guo G C 1998 {\it Phys. Rev. Lett.} {\bf 80} 4999\\
$[14]$   Duan L M and  Guo G C 1999 {\it Commun. Theor. Phys.} {\bf 31} 223\\
$[15]$   Galv\~{a}o E F and Hardy  L 2000 {\it Phys. Rev.} A {\bf 62} 022301\\
$[16]$  Gao T,  Yan F L, and  Wang Z X 2004 {\it J. Phys.} A: Math. Gen. {\bf 37} 3211\\
$[17]$   Gao T,  Yan F and  Wang Z, e-print archive quant-ph/0308036\\
$[18]$  Bechmann-Pasquinucci H and Gisin N 1999 {\it Phys. Rev.} A {\bf 59} 4238\\}
\end{document}